\documentclass[aps,prl,showpacs,showkeys,twocolumn]{revtex4-1}
\usepackage{graphicx,hyperref,amsmath,amsfonts,comment}
\usepackage{epstopdf,color,bm,multirow,rotating,soul}
\usepackage[T2A]{fontenc} \usepackage[utf8]{inputenc}
\usepackage[russian, english]{babel}
\begin{document}
\title[Noise-immunity Kazan quantum line at 143 km regular fiber link]{
Noise-immunity Kazan quantum line at 143 km regular fiber link}
\author{Oleg I. Bannik$^{1,2}$,
Lenar R. Gilyazov$^{1,2}$,
Artur V. Gleim$^{1}$,
Nikolay S. Perminov$^{1,2}$,
Konstantin S. Melnik$^{1,2}$,
Narkis M. Arslanov$^{1,2}$,
Aleksandr A. Litvinov$^{1,2}$, 
Albert R. Yafarov$^{1,2}$,
and
Sergey A. Moiseev$^{1,2,*}$}
\affiliation{$^{1}$ Kazan Quantum Center, Kazan National Research Technical University n.a. A.N.Tupolev-KAI, 10 K. Marx, Kazan 420111, Russia}
\affiliation{$^{2}$ Kazan Quantum Communication Ltd, 10 K. Marx, Kazan 420111, Russia}
\email{s.a.moiseev@kazanqc.org}
\begin{abstract}
We experimentally demonstrate a long-distance quantum communication at 143 km between the city of Kazan and the urban-type village of Apastovo in the Republic of Tatarstan by using quantum key distribution prototype providing high noise-immunity of network lines and nodes due to phase coding in subcarrier wave.
Average secret key generation rate was 12 bits per second with losses in the line of 37 decibels for a distance of 143 km during 16.5 hours of continuous field test.
The commercialization perspectives of the demonstrated long-range QKD system are discussed. 
\end{abstract}

\maketitle

\section{Intercity quantum networks}
One of the first commercial system of quantum key distribution (QKD)  using a standard fiber-optic channel in urban communication line between the Swiss cities of Geneva and Lausanne was implemented in the work \cite{stucki2002quantum}. 
The key distribution  was demonstrated over a distance of 67 km at a wavelength of 1550 nm with a quantum secret key rate (SKR) of about 60 bits per second (bps). 
This system used a phase-coded BB84 protocol \cite{bennett1984quantum,muller1997plug},
the main components of which were a polarization beamsplitter, phase modulator and single-photon detection module capable of operating at room temperature.
In most quantum communication schemes \cite{gisin2002quantum,andersen2015hybrid,diamanti2016practical,zhang2017quantum,yan2017quantum,acin2018quantum}, the expensive critical element is the photon detector, which largely determines the quantum signal utilization, signal to noise ratio, cost and reliability of the system, as well as its main functionality and complexity of maintenance during operation.
Currently, to ensure high quantum communication (QC) devices performance, two main types of quantum detectors are widely used: single-photon avalanche detector (SPAD) \cite{tosi2014low,zhang2015advances}  and high-efficiency superconducting single-photon detector (SSPD) \cite{natarajan2012superconducting,zhang2017nbn,zolotov2018photon,roy2018number,autebert2019direct}. 
Promising results for stable urban quantum networks and long-distance QCs have been recently obtained with SPAD and SSPD detectors.

\textit{QC with SPAD.}
The well-known demonstration of a commercial quantum communication system, described in work \cite{stucki2002quantum}, contained a SPAD detector composed of a cooled Peltier avalanche photodiode.
In \cite{dixon2015high}, 
by using phase-coding BB84 protocol \cite{scarani2008quantum,lo2005efficient} with Mach-Zehnder interferometers, UK-Japan group presented the prototype of a high bit rate QKD system providing a total of 878 Gbit of secure key data over a 34 day period that corresponded to a sustained key rate of around 300 kbps. 
The system was deployed over a standard 45 km link of an installed metropolitan telecommunication fiber network in central Tokyo. 
The prototype of QKD system is compact, robust and automatically stabilised, enabling key distribution during diverse weather conditions.
The security analysis has been performed for this system by taking into account finite key size effects, decoy states \cite{lucamarini2013efficient} with a quantified key failure probability of $\epsilon=10^{{-}10}$. 

Recently, Chinese group, known for creating a quantum satellite \cite{liao2017satellite}, in \cite{mao2018integrating} presented  integration of QKD on polarization coding decoy-state BB84 protocol \cite{chen2010metropolitan,yin2012quantum,liao2017satellite} with a commercial long-distance network of 3.6 Tbps classical data at 21 dBm launch power over 66 km fiber. 
This scheme is quite noise immunity in relation to the influences on devices in network nodes, since it does not contain interferometers, and is based on polarizing devices. 
With 20 GHz pass-band filtering and large effective core area fibers, real-time secure key rates can reach 4.5 kbps and 5.1 kbps for co-propagation and counter-propagation at the maximum launch power, respectively. This demonstrates feasibility and represents an important step towards building a quantum network that coexists with the current trunk fiber infrastructure of classical communications.
A significant increase in the maximum range of QKD systems is achieved by replacing SPAD detector with SSPD detector.

\textit{QC with SSPD.}
Ultra fast BB84 QKD transmission at 625 MHz clock rate through a 97 km field-installed fiber using practical clock synchronization based on wavelength-division multiplexing for time-bin coding protocol with Mach-Zehnder interferometers was demonstrated by Japan group in \cite{tanaka2008ultra}. 
The QC has been implemented  over-one-hour stable secret key generation at a 800 bps with low quantum bit error rate (QBER) of 2.9 $\%$ where the quantum information was additionally protected using decoy method \cite{lo2005decoy}.
These results open the way to global secure QCs capable of functioning at high losses in working optical lines \cite{hwang2003quantum}.

The experimental results for a long-term field trial of 1-GHz clock QKD were presented in  \cite{shimizu2013performance} by using the differential phase shift coding scheme \cite{molotkov2015analog} with Mach-Zehnder interferometers  incorporated in the test-bed optical fiber network installed in the Tokyo metropolitan area. 
A photon transmitter and a photon receiver were connected to the 90 km loop-back optical link with 30 dB loss. 
SSPD detectors were employed to detect photons with a low dark count rate. 
Stable maintenance-free operation was achieved for 25 days, where the average secure key generation rate 1.1 $\pm$ 0.5 kbps and the quantum bit error rate 2.6$\%$. 
The experiments have shown a significant impact of meteorological conditions on the main parameters of QKD system.

To move from existing commercial QKD-systems operating at distances of about 100 km to new solutions stable at ultra-long distances about of 150-200 km, it is desirable to use optical schemes that do not include interferometers.
In this work we realized stable interferometer-free one-way QKD system at 143 km regular fiber line by using  subcarrier waves phase-coding protocol and discuss prospective of this system for ultra-long distance QC.
\vspace{-0.5cm}

\section*{Noise immunity prototype for ultra-long distance QC}
\textit{Noise immunity problem.}
Noise immunity and active intelligent stabilization \cite{kulik2014method,kulik2014line,balygin2016active} are very important characteristics of systems for ultra-long distance QCs.
The characteristics allow to classify the existing long-distance QCs \cite{molotkov2008resistance,molotkov2011solution} by using extensive  quantum analysis \cite{molotkov2016complexity} (traditional for cryptography) and intelligent noise-analysis \cite{nigmatullin2017general,nigmatullin2019universal,nigmatullin2019detection,nigmatullin2016forecasting} (traditional for steganography, friend identification systems and general type complex systems).
For solution of the problems of noise immunity in relation to the influences on both the  fiber communication line and QKD apparatus in the network nodes, we must accordingly abandon the use of interferometers \cite{molotkov2004multiplex} reducing vibration resistance of nodes and polarization coding sensitive to deformations, vibration and temperature changes in the fiber.

\textit{Phase-coding as a solution.}
The unique noise immunity is inherent to the subcarrier wave phase-coding protocol
\cite{merolla1999single} what distinguishes this approach from many others \cite{wehner2018quantum,djordjevic2019recent,razavi2019quantum}.
A quantum signal is not directly generated by the source in the one-way QKD system of subcarrier frequencies, but is created as a result of phase modulation of carrier wave at Alice and Bob sides \cite{merolla1999single}. 
A description of a typical technological implementation of this protocol can be found in \cite{gleim2016secure}. 
Below we offer a modified variant of this QKD-system.

\begin{figure}[t]
\includegraphics[width = 0.48\textwidth]{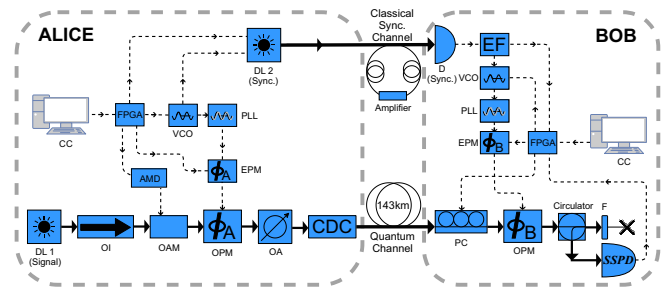}
\caption{ One-way scheme for the long-distance QC complex.
DL1 and DL2: diode laser 1 and 2; OI: optical isolator; OAM: optical amplitude modulator; OPM: optical phase modulator; OA: optical attenuator; FPGA: field-programmable gated-array and these serve (in conjunction with the control computers) as the control apparatus of this system; VCO: voltage control oscillator; PLL: phase looked loop; EPM: electrical phase modulator; AMD: amplitude modulator driver; CDC: chromatic dispersion compensator; PC: polarization controller; F: optical filter; SSPD: superconducting single-photon detector; EF: electrical filter; D: photodiode; $\Phi$: optical phase modulator; CC: classical computer.}
\label{scheme}
\vspace{-0.5cm}
\end{figure}

\textit{Components and properties.}
We propose a QKD-system of one-way QC shown in Fig. \ref{scheme}.
Tunable wavelength continuous wave mode DFB-laser with 5 kHz frequency linewidth was used as a carrier wavelength source for quantum channel. 
The wavelength was adjusted in accordance with the filter bandwidth in the receiving unit (Bob side). 
The laser radiation was launched to the electro-optical amplitude modulator for generation of short phase modulator was a sequence of 
2.5 ns pulses at the repetition rate of 100 MHz. 
Then the generated pulse were transmitted to the optical phase modulator (OPM) providing uniform phase modulation over the entire pulse duration.
4.8 GHz electromagnetic field was also applied to the radio-frequency input of OPM to generate optical sidebands.
The relative phase of the subcarrier waves is determined by the phase of the modulating electromagnetic field with modulation index of 0.033.
Further, the optical signal was attenuated by the variable attenuator to the sidebands power level defined by the protocol, namely, no more than 0.2 photons per modulation cycle. 
Chromatic dispersion compensator was introduced to eliminate the detrimental influence of chromatic dispersion effect on the interferometric visibility on the receiver unit side. Alice overall output power was 79.3 pW.
Both Alice’s and Bob’s phase modulators contained an linear polarizer aligned with the electro-optical crystal axis to ensure fully modulated signal. 

To maximize signal utilization, Bob used a polarization controller at its input to compensate for polarization distortions introduced by the fiber optics line. 
Following the maximum counting rate at the output of a single-photon detector, the controller automatically aligns the signal polarization along the axis of the OPM polarizer.
After the phase modulation, the input light was filtered at narrow band notch filter. 

Our approach uses only one sideband, introducing an additional 3 dB signal loss,
while that allows us to filter out both carrier wavelength and fiber line background spurious noise in a cheap simple way using only one spectral filter. 
One of the sidebands is launched to a SSPD-detector.
SSPD operates at a temperature of 2.1 K and has a low dark count rate of 0.5 Hz with a high quantum efficiency of 50 $\%$.

\section*{Experimental demonstration}
\begin{figure}[t]
\includegraphics[width = 0.48\textwidth]{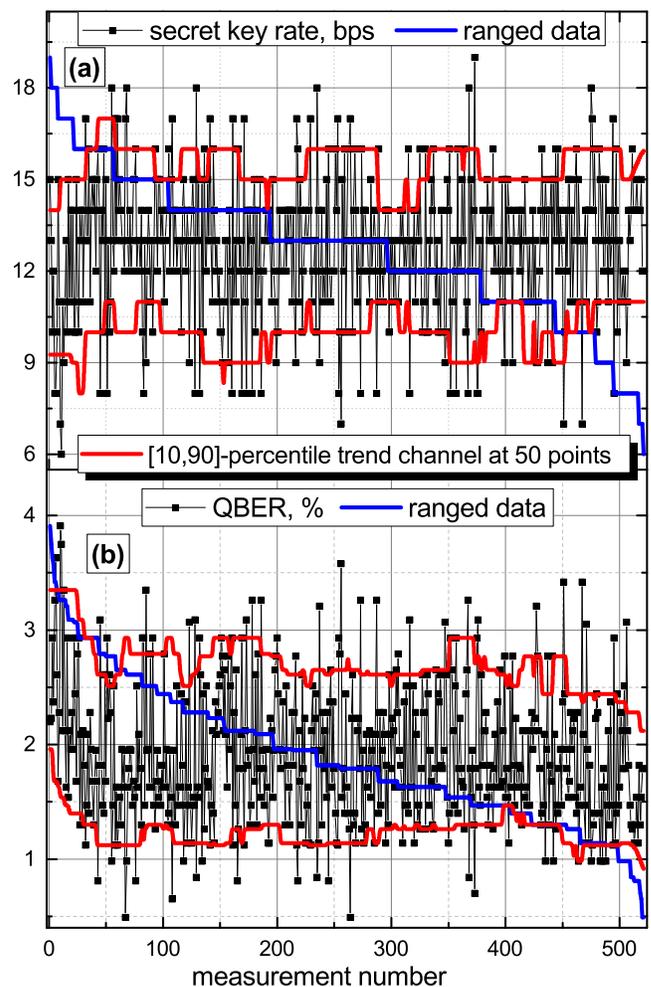}
\caption{Distributions for \textbf{(a)} secret key rate ($\sim 12$ bps) and \textbf{(b)} QBER ($\sim 2 \%$) for two-day measurements.}
\label{QBER}
\vspace{-0.5cm}
\end{figure}

For the field tests, we used Kazan-Apastovo regular fiber link with a length of 143 km and losses of 37 dB in the quantum channel and 45 dB in the synchronization channel with erbium-doped fiber amplifier.
The total time for continuous testing was 16.5 hours. 
About 700 kbit of key information was generated in the quantum channel with an average value of the secret quantum key generation rate of 12 bps.
The value of QBER was in the stable range from 0.5 $\%$ to 3.5 $\%$, on average, the value of QBER $\sim$ 2 $\%$ (see Fig. \ref{QBER}). 
This generation rate allows to change the 256-bit encryption key up to two times per minute.
The data of the stable experimental demonstration show the possible using of the proposed QKD system in commercial long-range QCs.

In our work, the results were obtained in a long-distance and regular trunk telecommunication fiber between the cities of the Russian Federation.
Basic characteristics of the tests of our QKD system for urban quantum networks \cite{bannik2017multinode,news1} and intercity trunk lines \cite{news2} are presented in Tab. \ref{T:param}. For comparison, Tab. \ref{T:param} also presents the basic characteristics of the well-known long-range QKD systems discussed in the introduction which were also tested under real conditions \cite{notes1}. 

\begin{table}[ph]
\begin{tabular}{|c|c|c|c|c|}
\hline
\multicolumn{5}{|c|}{Kazan quantum lines} \\ \hline
Length& SKR & Detector & Loss (dB); & Group \\
(km) & (bps) &  & QBER (\%) &  \\ \hline
12 & $2 \cdot10^4$ & $^*$SPAD & 7; 4 & Russia \\ \hline
143 & 12 & $^\dagger$SSPD & 37; 2 & Russia \\ \hline
\multicolumn{5}{|c|}{International commercial systems} \\ \hline
67 & 60 & \cite{stucki2002quantum} SPAD & 14; 6 & Switzerland \\ \hline
45 & $3 \cdot10^5$ & \cite{dixon2015high} SPAD & 14; 4 & Japan \\ \hline
66 & $5 \cdot10^5$ & \cite{mao2018integrating} SPAD & 21; 5 & China \\ \hline
97 & 800 & \cite{tanaka2008ultra} SSPD & 33; 3 & Japan \\ \hline
90 & $1 \cdot10^3$ & \cite{shimizu2013performance} SSPD & 30; 3 & Japan \\ \hline
\end{tabular}
\caption{Basic characteristics of the proposed QKD system: $^*$Kazan city quantum network (previous field tests, Kazan, May 2017 \cite{bannik2017multinode,news1}), $^\dagger$intercity long-distance QC (current experiment, Kazan-Apastovo, August 2019 \cite{news2}).}
\label{T:param}
\vspace{0cm}
\end{table}

\noindent
It is worth noting that in \cite{bacco2019field,choi2014field} and \cite{walenta2015towards,boaron2018secure} tests of various QKD systems at shorter distances and using ultra-low loss fibers are presented.

\section{Innovation outlook}
Based on the demonstrated QKD prototype, we propose the concept of a universal QC complex $"$Kazan-Q1$"$ (see outlook Tab. \ref{T:complex} and patents \cite{patent1,patent2,patent3,patent4,patent5}) which combines the technological simple solutions, noise immunity of the subcarrier wave coding and robust quantum-classical information protection based on the extended statistical data mining against new types of quantum attacks \cite{bykovsky2018quantum}.

\begin{table}[ht]
\begin{tabular}{|c|c||c|c|}
\hline
Name & Kazan-Q1 & Detector & SSPD \\
 &  &  &  or SPAD \\ \hline
Coding & Phase in & Active & Phase \\
 & subcarrier wave & element & modulator \\ \hline
Security & Quantum and & QRNG & On PD with \\
 & classical & & robust-defense \\ \hline
Noise & Network nodes & Decoys and & Amplitude \\
immunity & and lines & diagnostics & modulator and AI \\ \hline 
\end{tabular}
\caption{Outlook for the complex $"$Kazan-Q1$"$. Here: QRNG: quantum random number generator; PD: PIN-photodiode; AI: artificial intelligence.}
\label{T:complex}
\vspace{-0.3cm}
\end{table}
\noindent
We see the following distinctive features and prospects for the development of the long-distance QC complex $"$Kazan-Q1$"$ in the context of world achievements in the area of innovative quantum technologies:

1) statistical monitoring with AI for keys and electronic components based on the robust nonparametric criteria \cite{nigmatullin2010new,perminov2018rcf,nigmatullin2019universal,smirnov2018sequences,perminov2017sra};

2) compact \cite{melnik2018using,zhang2019integrated,eriksson2019wavelength} planar implementation \cite{orieux2016recent} without using interferometers for the main components \cite{acin2018quantum}:
beam splitter  \cite{desiatov2019ultra,tao2008cascade,thomaschewski2019plasmonic}, planar phase modulator \cite{ren2019integrated}, high purity QRNG on PD \cite{campbell2016recent} with robust-defense \cite{perminov2018rcf,avesani2018source,ioannou2018much,grangier2018quantum,drahi2019certified};

3) implementation of advanced decoy states \cite{molotkov2019there,huang2018quantum} in a complex using intelligent monitoring methods \cite{perminov2018rcf,nigmatullin2019universal} for quantum diagnostics of network integrity and intrusion detection
\cite{kravtsov2018relativistic,gaidash2019countermeasures,gaidash2019methods};

4) advanced post-processing \cite{jia2019quantum} for error correction and modeling in optical simulator for testing and specialization of software for specific urban conditions.

\textit{Acknowledgments.}
Research is financially supported by a grant of the Government of the Russian Federation, project No. 14.Z50.31.0040, February 17, 2017.

We are very grateful to the companies of PJSC $"$Tattelecom$"$ and PJSC $"$Rostelecom$"$ for providing working fiber-optic communication lines and for officially documenting Russia's record \cite{news2} for the range of long-distance QC during this experimental demonstration.

\section{Information about authors}
\noindent
\textbf{Oleg Igorevich Bannik},\\
(b. 1988), in 2012 graduated from the faculty of electronics of Saint Petersburg Electrotechnical University in the direction of "Electronics and Microelectronics", researcher at the Kazan Quantum Center of the KNRTU-KAI.\\
\textit{Area of interest:} quantum communications, optoelectronics, photonics.\\
\textit{E-mail:} olegbannik@gmail.com\\
\\
\noindent
\textbf{Lenar Rishatovich Gilyazov},\\
(b. 1985), in 2008 graduated from the the Physics Department of Kazan Federal University, researcher at the Kazan Quantum Center of the KNRTU-KAI.\\
\textit{Area of interest:} quantum communications, optoelectronics, photonics.\\
\textit{E-mail:} lgilyazo@mail.ru\\
\\
\noindent
\textbf{Artur Viktorovich Gleim},\\
(b. 1990), in 2012 graduated from ITMO University in the direction of "Photonics and Optoinformatics", candidate of technical sciences, head of lab. at the Kazan Quantum Center of the KNRTU-KAI.\\
\textit{Area of interest:} quantum communications, optoelectronics, photonics.\\
\textit{E-mail:} aglejm@yandex.ru\\
\\
\noindent
\textbf{Nikolay Sergeevich Perminov},\\
(b. 1985), in 2008 graduated the department of Theory of Relativity and Gravity of the Physics Department of Kazan Federal University in the direction of “Physics”, researcher at the Kazan Quantum Center of the KNRTU-KAI.\\
\textit{Area of interest:} optimal control, quantum informatics, statistics, software, economics.\\
\textit{E-mail:} qm.kzn@ya.ru\\
\\
\noindent
\textbf{Konstantin Sergeevich Melnik},\\
(b. 1993), in 2018 graduated from Institute of Radio Electronics and Telecommunications of
Kazan National Research Technical University in the direction of "Radio Engineering", PhD student  at the Kazan Quantum Center of the KNRTU-KAI.\\
\textit{Area of interest:} quantum communications, optoelectronics, photonics.\\
\textit{E-mail:} mkostyk93@mail.ru\\
\\
\noindent
\textbf{Narkis Musavirovich Arslanov},\\
(b. 1980), in 2003 graduated from the the Physics Department of Kazan Federal University, senior researcher at the Kazan Quantum Center of the KNRTU-KAI.\\
\textit{Area of interest:} quantum communications, photonics.\\
\textit{E-mail:} narkis@yandex.ru\\
\\
\noindent
\textbf{Aleksandr Alekseevich Litvinov},\\
(b. 1985), in 2008 graduated the department of Theory of Relativity and Gravity of the Physics Department of Kazan Federal University in the direction of “Physics”, engineer at the Kazan Quantum Center of the KNRTU-KAI.\\
\textit{Area of interest:} quantum communications, optoelectronics, robust methods, quantum memory, software, economics.\\
\textit{E-mail:} litvinov85@gmail.com\\
\\
\noindent
\textbf{Albert Ruslanovich Yafarov},\\
(b. 1977), in 2006 graduated from the Department of Geography of Kazan State University in 2006 with a degree in Geography, engineer at the Kazan Quantum Center of the KNRTU-KAI.\\
\textit{Area of interest:} quantum communications, robust methods, quantum memory, project management, public relations, economics.\\
\textit{E-mail:} a.r.yafarov@gmail.com\\
\\
\noindent
\textbf{Sergey Andreevich Moiseev},\\
(b. 1957), in 1979 graduated with honors from the physical faculty of Kazan State University in the direction of "Radiophysics". Doctor of Phys.-Math. Sciences, prof. Department of Radio-photonics and Microwave Technologies, KNRTU-KAI, Director of the Kazan Quantum Center KNRTU-KAI.\\
\textit{Area of interest:} optics and spectroscopy, quantum memory, quantum informatics, quantum communications, nonlinear and coherent optics.\\
\textit{E-mail:} s.a.moiseev@kazanqc.org\\

\vspace{-1.0cm}

\bibliographystyle{ieeetr}
\bibliography{LD_QC}

\end{document}